\documentclass[10pt,conference]{IEEEtran} 
\usepackage{cite,graphicx,psfrag,amssymb,amsmath,url,color}
\newtheorem{theorem}{Theorem}
\newtheorem{lemma}{Lemma}

\def\definedas{\triangleq}
\def\opt{{\mathop{\rm opt}}}
\def\root{{\mathop{\rm root}}}
\def\boldl{{\mbox{\boldmath $l$}}}
\def\boldsl{{\mbox{\scriptsize \boldmath $l$}}}

\def\Cost{L}
\def\lu{\lambda}
\def\q{a}

\def\L{{{\mathcal L}_n}}
\def\X{{\mathcal X}}
\def\Z{{\mathbb Z}}

\begin{document}
\title{Tight Bounds on Minimum Maximum Pointwise Redundancy}
\bibliographystyle{IEEEtran}

\author{\authorblockN{Michael B. Baer}
\authorblockA{vLnks\\
Mountain View, CA  94041-2803, USA\\
Email:{\color{white}{.}}calbear{\color{black}{@}}{\bf \scriptsize \.{1}}eee.org}}

\maketitle

\begin{abstract}
This paper presents new lower and upper bounds for the optimal compression of binary prefix codes in terms of the most probable input symbol, where compression efficiency is determined by the nonlinear codeword length objective of minimizing maximum pointwise redundancy.  This objective relates to both universal modeling and Shannon coding, and these bounds are tight throughout the interval.  The upper bounds also apply to a related objective, that of $d$\textsuperscript{th} exponential redundancy.
\end{abstract}

\section{Introduction} 

A lossless binary prefix coding problem takes a probability mass
function $p(i)$, defined for all $i$ in the input alphabet
$\X$, and finds a binary code for~$\X$.  Without loss of generality,
we consider an $n$-item source emitting symbols drawn from the
alphabet $\X = \{1, 2, \ldots, n\}$ where $\{p(i)\}$ is the sequence of
probabilities for possible symbols ($p(i) > 0$ for $i \in \X$
and $\sum_{i \in \X} p(i) = 1$) in monotonically nonincreasing order ($p(i)
\geq p(j)$ for $i<j$).  The source symbols are coded into binary
codewords.  The codeword $c(i) \in \{0,1\}^*$ in code~$c$,
corresponding to input symbol~$i$, has length $l(i)$, defining length
vector~$\boldl$.

The goal of the traditional coding problem is to find a prefix
code minimizing expected codeword length $\sum_{i \in \X}
p(i) l(i)$, or, equivalently, minimizing average redundancy
$$\bar{R}(\boldl,p) \definedas \sum_{i \in \X} p(i) l(i) - H(p) =
\sum_{i \in \X} p(i) (l(i)+\lg p(i))$$ where $H$ is $-\sum _{i \in \X}
p(i)\lg p(i)$, Shannon entropy, and $\lg \definedas \log_2$.  A prefix
code is a code for which no codeword begins with a sequence that also
comprises the whole of a second codeword.  This problem is equivalent
to finding a minimum-weight external path
$$\sum_{i \in \X} w(i) l(i)$$ among all rooted binary trees, due to
the fact that every prefix code can be represented as a binary tree.
In this tree representation, each edge from a parent node to a child
node is labeled $0$ (left) or $1$ (right), with at most one of each type of edge per parent node.  A leaf is a node without children; this
corresponds to a codeword, and the codeword is determined by the path
from the root to the leaf.  Thus, for example, a leaf that is the
right-edge ($1$) child of a left-edge ($0$) child of a left-edge ($0$) child of the
root will correspond to codeword~$001$.  Leaf depth (distance from the
root) is thus codeword length.  The weights are the probabilities
(i.e., $w(i) = p(i)$), and, in fact, we will refer to the problem
inputs as $\{w(i)\}$ for certain generalizations in which their sum,
$\sum_{i \in \X} w(i)$, need not be~$1$.

If formulated in terms of $\boldl$, the constraints on the
minimization are the integer constraint (i.e., that codes must be of integer
length) and the Kraft inequality\cite{McMi}; that is, the set of
allowable codeword length vectors is
$$\L \definedas \left\{\boldl \in \Z_+^n \mbox{ such that }
\sum_{i=1}^n 2^{-l(i)} \leq 1\right\}.$$ 

Drmota and Szpankowski\cite{DrSz} investigated a problem which,
instead of minimizing average redundancy $\bar{R}(\boldl,p) \definedas
\sum_{i \in \X} p(i) (l(i)+\lg p(i))$, minimizes maximum pointwise
redundancy
$$R^*(\boldl,p) \definedas \max_{i \in \X} (l(i)+\lg p(i)).$$ Related
to a universal modeling problem \cite[p.~176]{Shta}, the idea here is
that, given a symbol to be compressed, we wish the length of the
compressed data ($l(i)$) to exceed self-information ($-\lg p(i)$) by
as little as possible, and thus consider the worst case in this
regard.  This naturally relates to Shannon coding, as a code with
lengths $\lceil -\lg p(i) \rceil$ would never exceed self-information
by more than $1$ bit.  Any solution, then, would necessarily have no
codeword longer than its Shannon code counterpart.  Indeed, Drmota and
Szpankowski used a generalization of Shannon coding to solve the
problem, which satisfies $$0 \leq R^*(\boldl^\opt,p) < 1.$$
We will improve the bounds, given $p(1)$, for minimum maximum
pointwise redundancy and discuss the related issue of the length of
the most likely codeword in these coding problems.  These bounds are
the first of their kind for this objective, analogous to those for
traditional Huffman coding\cite{Gall,John,CGT,MoAb,CaDe1,Mans} and
other nonlinear codes\cite{Tane,BlMc,BaerI06}.

The bounds are derived using an alternative solution to this
problem, a variation of Huffman coding\cite{Baer05} derived from that
in \cite{Golu}.  In order to explain this variation, we first review
the Huffman algorithm and some of the ways in which it can be
modified.

It is well known that the Huffman algorithm\cite{Huff} finds a code
minimizing average redundancy.  The Huffman algorithm is a greedy
algorithm built on the observation that the two least likely symbols
will have the same length and can thus be considered siblings in the
coding tree.  A reduction can thus be made in which the two symbols
with weights $w(i)$ and $w(j)$ can be considered as one with combined
weight $w(i)+w(j)$, and the codeword of the combined item determines
all but the last bit of each of the items combined, which are
differentiated by this last bit.  This reduction continues until there
is one item left, and, assigning this item the null string, a code is
defined for all input symbols.  In the corresponding optimal code
tree, the $i$\textsuperscript{th} leaf corresponds to the codeword of
the $i$\textsuperscript{th} input item, and thus has weight $w(i)$,
whereas the weight of parent nodes are determined by the combined
weight of the corresponding merged item.  Van Leeuwen gave an
implementation of the Huffman algorithm that can be accomplished in
linear time given sorted probabilities\cite{Leeu}.  Shannon\cite{Shan}
had previously shown that an optimal $\boldl^\opt$ must satisfy
$$H(p) \leq \sum_{i \in \X} p(i)l^\opt(i) < H(p) + 1 \mbox{, i.e., }
0 \leq \bar{R}(\boldl^\opt,p) < 1.$$
 
Simple changes to the Huffman algorithm solve several related coding
problems which optimize for different objectives.  Generalized
versions of the Huffman algorithm have been considered by many
authors\cite{HKT,Park,Knu1,ChTh}.  These generalizations change the
combining rule; instead of replacing items $i$ and $j$ with an item of
weight $w(i)+w(j)$, the generalized algorithm replaces them with an
item of weight $f(w(i),w(j))$ for some function~$f$.  Thus the weight
of a combined item (a node) no longer need be equal to the sum of the
probabilities of the items merged to create it (the sum of the leaves
of the corresponding subtree).  This has the result that the sum of
weights in a reduced problem need not be~$1$, unlike in the original
Huffman algorithm.  In particular, the weight of the root, $w_\root$,
need not be~$1$.  However, we continue to assume that the sum of
$p(\cdot)$, the inputs before reduction, will always be~$1$.

One such variation of the Huffman algorithm was used in Humblet's 
dissertation\cite{Humb0} for a queueing application (and further
discussed in \cite{HKT, Park, Humb2}).  The problem this variation
solves is as follows: Given probability mass function $p$ and $a>1$,
find a code minimizing
\begin{equation} 
\Cost_\q(p,\boldl) \definedas \log_\q \sum_{i \in \X} p(i) \q^{l(i)} .
\label{one} 
\end{equation} 
This growing exponential average problem is solved by using combining rule 
\begin{equation}
f(w(i),w(j)) = \q w(i)+\q w(j).
\label{expcomb} 
\end{equation}  
This problem was proposed (without solution) by Campbell\cite{Camp0},
who later noted that this formulation can be extended to decaying
exponential base $\q \in (0,1)$\cite{Camp}; Humblet noted that the
Huffman combining method (\ref{expcomb}) finds the optimal code for
(\ref{one}) with $\q \in (0,1)$ as well\cite{Humb2}.

Another variation, proposed in \cite{Nath} and
solved for in \cite{Park}, can be called $d$\textsuperscript{th} exponential
redundancy\cite{Baer05}, and is the minimization of the following:
$$R^d(\boldl,p) \definedas \frac{1}{d} \lg \sum_{i \in \X} p(i)^{1+d}
2^{dl(i)}.$$  Here we assume that $d>0$, although $d \in (-1,0)$ is
also a valid problem.  Clearly, this can be solved via reduction to
(\ref{one}) by assigning $\q = \lg d$ and using input weights $w(i) =
p(i)^{1+d}$.

Minimizing maximum redundancy is equivalent to minimizing
$d$\textsuperscript{th} exponential redundancy for $d \rightarrow
\infty$.  This observation leads to a Huffman-like solution with the
combination rule
\begin{equation}
f(w(i),w(j))=2\max(w(i),w(j))
\label{mmprcomb}
\end{equation}  
as in \cite{Baer05}.

In the next section, we find tight exhaustive bounds for
the values of optimal $R^*(\boldl,p)$ and corresponding $l(1)$ in
terms of~$p(1)$, then find how we can extend these to 
exhaustive --- but not tight --- bounds for optimal $R^d(\boldl,p)$.  

\section{Bounds on the Redundancy Problems}
\label{bred}

It is useful to come up with bounds on the performance of an optimal
code, often in terms of the most probable symbol, $p(1)$.  In minimizing average redundancy, such bounds are often referred to as
``redundancy bounds'' because they are in terms of this average redundancy,
$\bar{R}(\boldl,p) = \sum_{i \in \X} p(i) l(i) - H(p)$.  The simplest
bounds for the optimal solution to the minimum maximum pointwise
redundancy problem
$$R_\opt^*(p) \definedas \min_{\boldsl \in \L} \max_{i \in \X}
\left(l(i)+\lg p(i)\right)$$ can be combined
with those for the average redundancy problem:
\begin{equation}
0 \leq \bar{R}_\opt(p) \leq R_\opt^*(p) < 1
\label{mmprbounds}
\end{equation}
where $\bar{R}_\opt(p)$ is the average redundancy of the average
redundancy-optimal code.  The average redundancy case is a lower bound
because the maximum ($R^*(\boldl,p)$) of the values ($l(i)+\lg p(i)$)
that average to a quantity ($\bar{R}(\boldl,p)$) can be no less than
the average (a fact that holds for all $\boldl$ and~$p$).  The upper
bound is found similarly to the average redundancy case; we can note
that Shannon code $l_p^0(i) \definedas \lceil -\lg p(i) \rceil$
results in $R_\opt^*(p) \leq R^*(\boldl_p^0,p) = \max_{i \in
\X}{(\lceil -\lg p(i) \rceil + \lg p(i))} < 1$.

A few observations can be used to find a series of improved lower and
upper bounds on optimum maximum pointwise redundancy based on
(\ref{mmprbounds}):

\begin{lemma}
Suppose we apply (\ref{mmprcomb}) to find a Huffman-like code tree in
order to minimize maximum pointwise redundancy.  Then the following
holds:
\begin{enumerate}
\item Items are always merged by nondecreasing weight.
\item The weight of the root $w_\root$ of the coding tree
determines the maximum pointwise redundancy, $R^*(\boldl,p) = \lg
w_\root$.
\item The total probability of any subtree is no greater than the
total weight of the subtree.  
\item If $p(1) \leq 2p(n-1)$, then a minimum maximum pointwise
redundancy code can be represented by a complete tree, that is, a tree
with leaves at depth $\lfloor \lg n \rfloor$ and $\lceil \lg n \rceil$
only (with $\sum_{i \in \X} 2^{-l(i)} = 1$). 
\end{enumerate}
\label{complete}
\end{lemma}

\begin{proof}
We use an inductive proof in which base cases of sizes $1$ and $2$ are
trivial, and we use weights~$w$, instead of probabilities $p$, to
emphasize that the sums of weights need not necessarily add up to~$1$.
Assume first that all properties here are true for trees of size $n-1$
and smaller.  We wish to show that they are true for trees of size~$n$.

The first property is true because
$f(w(i),w(j))=2\max(w(i),w(j))>w(i)$ for any $i$ and~$j$; that is, a
compound item always has greater weight than either of the items
combined to form it.  Thus, after the first two weights are combined,
all remaining weights, including the compound weight, are no less than
either of the two original weights.

Consider the second property; after merging the two least weighted of
$n$ (possibly merged) items, the property holds for the resulting
$n-1$ items.  For the $n-2$ untouched items, $l(i)+\lg w(i)$ remains
the same.  For the two merged items, let $l(n-1)$ and $w(n-1)$ denote
the maximum depth/weight pair for item $n-1$ and $l(n)$ and $w(n)$ the
pair for $n$.  If $l'$ and $w'$ denote the depth/weight pair of the
combined item, then $l'+\lg w' = l(n) - 1 + \lg (2 \max(w(n-1),w(n)))
= \max(l(n-1) + \lg w(n-1),l(n) + \lg w(n))$, so the two trees have
identical maximum redundancy, which is equal to $\lg w_\root$ since
the root node is of depth~$0$.  Consider, for example, $p =
(0.5,0.3,0.2)$, which has optimal codewords with lengths $\boldl =
(1,2,2)$.  The first combined pair has $l'+\lg w' = 1+\lg 0.6 =
\max(2+\lg 0.3,2+\lg 0.2) = \max(l(2)+\lg p(2),l(3)+\lg p(3))$.  This
value is identical to that of the maximum redundancy, $\lg 1.2 = \lg
w_\root$.

For the third property, the first combined pair yields a weight that
is no less than the combined probabilities.  Thus, via induction, the
total probability of any (sub)tree is no greater than the weight of
the (sub)tree.

In order to show the final property, first note that $\sum_{i \in \X}
2^{-l(i)} = 1$ for any tree created using the Huffman-like procedure,
since all internal nodes have two children.  Now think of the
procedure as starting with a queue of input items, ordered by
nondecreasing weight from head to tail.  After merging two items,
obtained from the head of the queue, into one compound item, that item
is placed back into the queue as one item, but not necessarily at the
tail; an item is placed such that its weight is no smaller than any
item ahead of it and is smaller than any item behind it.  In keeping
items ordered, this results in an optimal coding tree.  A variant of
this method can be used for linear-time coding\cite{Baer05}.

In this case, we show not only that an optimal complete tree exists,
but that, given an $n$-item tree, all items that finish at level
$\lceil \lg n \rceil$ appear closer to the head of the queue than any
item at level $\lceil \lg n \rceil - 1$ (if any), using a similar
approach to the proof of Lemma~2 in \cite{Baer07}.  Suppose this is
true for every case with $n-1$ items for $n>2$, that is, that all
nodes are at levels $\lfloor \lg (n-1) \rfloor$ or $\lceil \lg (n-1)
\rceil$, with the latter items closer to the head of the queue than
the former.  Consider now a case with $n$ nodes.  The first step of
coding is to merge two nodes, resulting in a combined item that is
placed at the end of the combined-item queue, as we have asserted that
$p(1) \leq 2p(n-1) = 2\max(p(n-1),p(n))$.  Because it is at the end
of the queue in the $n-1$ case, this combined node is at level
$\lfloor \lg (n-1) \rfloor$ in the final tree, and its children are at
level $1+\lfloor \lg (n-1) \rfloor = \lceil \lg n \rceil$.  If $n$ is
a power of two, the remaining items end up on level $\lg n = \lceil
\lg (n-1) \rceil$, satisfying this lemma.  If $n-1$ is a power of two,
they end up on level $\lg (n-1) = \lfloor \lg n \rfloor$, also
satisfying the lemma.  Otherwise, there is at least one item ending up
at level $\lceil \lg n \rceil = \lceil \lg (n-1) \rceil$ near the head
of the queue, followed by the remaining items, which end up at level
$\lfloor \lg n \rfloor = \lfloor \lg (n-1) \rfloor$.  In any case, all
properties of the lemma are satisfied for $n$ items, and thus for any
number of items.
\end{proof}

We can now present the improved redundancy bounds.

\begin{theorem}
For any distribution in which $p(1) \geq 2/3$, $R_\opt^*(p) = 1+\lg
p(1)$.  If $p(1) \in [0.5,2/3)$, then $R_\opt^*(p) \in [1+\lg
p(1),2+\lg (1-p(1)))$ and these bounds are tight.  Define $\lu
\definedas \lceil - \lg p(1) \rceil$, which, for $p(1) \in (0,0.5)$, is
greater than~$1$.  For this range the following bounds for
$R_\opt^*(p)$ are tight:
$$
\begin{array}{ll}
\quad p(1) & \quad R_\opt^*(p) \\
\hline  
& \\[-5pt]
\left[\frac{1}{2^{\lu}},\frac{1}{2^{\lu}-1}\right) &
\left[\lu+\lg p(1),1+\lg \frac{1-p(1)}{1-2^{-\lu}}\right) \\[6pt]
\left[\frac{1}{2^{\lu}-1},\frac{2}{2^{\lu}+1}\right) &
\left[\lg \frac{1-p(1)}{1-2^{-\lu+1}},1+\lg \frac{1-p(1)}{1-2^{-\lu}}\right) \\[6pt]
\left[\frac{2}{2^{\lu}+1},\frac{1}{2^{\lu-1}}\right) &
\left[\lg \frac{1-p(1)}{1-2^{-\lu+1}},\lu+\lg p(1) \right]\\
~&~\\
\end{array}
$$
\label{mmprbetter}
\end{theorem}

\begin{proof}
The key here is generalizing the simple bounds of (\ref{mmprbounds}).

${}$

\textit{Upper bound}:
Let us define what we call a \textit{first-order Shannon code}:
$$
l_p^1(i) = \left\{
\begin{array}{ll}
\lu \definedas \left\lceil -\lg p(1) \right\rceil,& i = 1 \\ 
\left\lceil -\lg
\left(p(i)\left(\frac{1-2^{-\lu}}{1-p(1)}\right)\right) \right\rceil,& i \in \{2, 3, \ldots, n\}
\end{array}
\right.
$$ This code, previously presented in the context of finding
\textit{average} redundancy bounds given \textit{any} probability
\cite{YeYe2}, improves upon the original ``zero-order'' Shannon code
$l_p^0$ by taking the length of the first codeword into account when
designing the rest of the code.  The code satisfies the Kraft
inequality, and thus, as a valid code, its redundancy is an upper
bound on the redundancy of an optimal code.  Note that
$$
\begin{array}{l}
\max_{i > 1} (l_p^1(i) + \lg p(i)) \\
\qquad \qquad = \max_{i>1} \left(
\left\lceil \lg 
\frac{1-p(1)}{p(i)(1-2^{-\lu})} \right\rceil
+ \lg p(i)\right) \\
\qquad \qquad < 1+\lg \frac{1-p(1)}{1-2^{-\lu}}.
\end{array}
$$
If $p(1) \in [2/(2^{\lu}+1),1/2^{\lu-1})$, the maximum pointwise redundancy of
the first item is no less than $1+\lg ((1-p(1))/(1-2^{-\lu}))$, and thus
$R_\opt^*(p) \leq R^*(\boldl_p^1,p) = \lu+\lg p(1)$.  Otherwise,
$R_\opt^*(p) \leq R^*(\boldl_p^1,p) < 1+\lg ((1-p(1))/(1-2^{-\lu}))$.

The tightness of the upper bound in $[0.5,1)$ is shown via
$$p = \left(p(1),1-p(1)-\epsilon,\epsilon\right)$$ for which the bound
is achieved in $[2/3,1)$ for any $\epsilon \in (0,(1-p(1))/2]$ and
approached in $[0.5,2/3)$ as $\epsilon \downarrow 0$.  If $\lu > 1$
and $p(1) \in [2/(2^{\lu}+1),1/2^{\lu-1})$, use probability mass
function
$$p = \left(p(1),\underbrace{\frac{1-p(1)-\epsilon}{2^\lu-2},
\ldots,\frac{1-p(1)-\epsilon}{2^\lu-2}}_{2^\lu-2}, \epsilon\right)$$
where $$\epsilon \in (0, 1-p(1)2^{\lu-1}).$$
Because $p(1) \geq 2/(2^{\lu}+1)$, $1-p(1)2^{\lu-1} \leq
(1-p(1)-\epsilon)/(2^\lu-2)$, and $p(n-1) \geq p(n)$.  Similarly,
$p(1) < 1/2^{\lu-1}$ assures that $p(1) \geq p(2)$, so the
probability mass function is monotonic.  Since $2p(n-1) > p(1)$, 
by Lemma~\ref{complete}, an optimal code for this probability mass
function is $l(i) = \lu$ for all~$i$, achieving $R^*(\boldl,p) = \lu
+ \lg p(1)$, with item $1$ having the maximum pointwise redundancy.

This leaves only $p(1) \in [1/2^{\lu}, 2/(2^{\lu}+1))$, for which we
consider $$p = \left(p(1),\underbrace{\frac{1-p(1)-\epsilon}{2^\lu-1},
\ldots,\frac{1-p(1)-\epsilon}{2^\lu-1}}_{2^\lu-1}, \epsilon\right)$$
where $\epsilon \downarrow 0$.  This is a monotonic probability mass
function for sufficiently small~$\epsilon$, for which we also have
$p(1) < 2p(n-1)$, so (again from Lemma~\ref{complete}) this results in
optimal code where $l(i) = \lu$ for $i \in \{1, 2, \ldots, n-2\}$ and
$l(n-1)=l(n)=\lu+1$, and thus the bound is approached with item $n-1$
having the maximum pointwise redundancy.

${}$

\textit{Lower bound}: Consider all optimal codes with $l(1) = \mu$ for
some fixed $\mu \in \{1, 2, \ldots\}$.  If $p(1) \geq 2^{-\mu}$,
$R^*(\boldl,p) \geq l(1) + \lg p(1) = \mu + \lg p(1)$.  If $p(1) <
2^{-\mu}$, consider the weights at level $\mu$ (i.e., $\mu$ edges
below the root).  One of these weights is $p(1)$, while the rest are
known to sum to a number no less than $1-p(1)$.  Thus at least one
weight must be at least $(1-p(1))/(2^{\mu}-1)$ and $R^*(\boldl,p) \geq
\mu + \lg ((1-p(1))/(2^{\mu}-1))$.  Thus,
$$R_\opt^*(p) \geq \mu + \lg \max\left(p(1),
\frac{1-p(1)}{2^{\mu}-1}\right)$$ for $l(1) = \mu$, and, since $\mu$
can be any positive integer,
$$R_\opt^*(p) \geq \min_{\mu \in \{1,2,3,\ldots\}} \left(\mu + \lg
\max\left(p(1), \frac{1-p(1)}{2^{\mu}-1}\right)\right)$$ which is equivalent
to the bounds provided.

For $p(1) \in [1/(2^{\mu+1}-1),1/2^{\mu})$ for some $\mu$, consider
$$\left(p(1), \underbrace{\frac{1-p(1)}{2^{\mu+1}-2}, \ldots,
\frac{1-p(1)}{2^{\mu+1}-2}}_{2^{\mu+1}-2}\right).$$ By
Lemma~\ref{complete}, this will have a complete coding tree and thus
achieve the lower bound for this range ($\lu=\mu+1$).  Similarly
$$\left(p(1), \underbrace{2^{-\mu-1}, \ldots,
2^{-\mu-1}}_{2^{\mu+1}-2}, 2^{-\mu}-p(1)\right)$$ has a
fixed-length optimal coding tree for $p(1) \in [1/2^\mu,
1/(2^\mu-1))$, achieving the lower bound for this range ($\lu=\mu$).
\end{proof}

\begin{figure}[t]
     \psfrag{R}{\Large $R_\opt^*(p)$}
     \psfrag{p(1)}{\Large $p(1)$}
     \centering
          \resizebox{8.75cm}{!}{\includegraphics{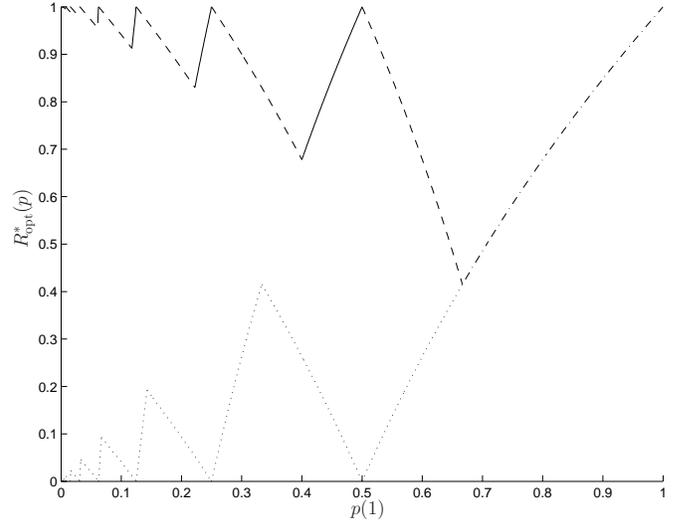}}
     \caption{Tight bounds on minimum maximum pointwise redundancy,
     including achievable upper bounds (solid), approachable upper
     bounds (dashed), achievable lower bounds (dotted), and fully
     determined values for $p(1)\geq 2/3$ (dot-dashed).}
     \label{mmprComplete}
\end{figure}

Note that the bounds of (\ref{mmprbounds}) are identical to the tight
bounds at powers of two.  In addition, the tight bounds clearly
approach $0$ and $1$ as $p(1) \downarrow 0$.  This behavior is in
stark contrast with average redundancy, for which bounds
get closer, not further apart, due to Gallager's redundancy
bound\cite{Gall} --- $\bar{R}_\opt(p) \leq p(1)+0.086$ --- which
cannot be significantly improved for small $p(1)$\cite{Mans}.
Moreover, approaching $1$, the upper and lower bounds on minimum average redundancy
coding converge but never merge, whereas the minimum maximum
redundancy bounds are identical for $p(1) \geq 2/3$.

In addition to finding redundancy bounds in terms of $p(1)$, it is
also often useful to find bounds on the behavior of $l(1)$ in terms of
$p(1)$, as was done for optimal average redundancy in \cite{CaDe2}.

\begin{theorem}
Any optimal code for probability mass function $p$, where $p(1) \geq
2^{-\nu}$, must have $l(1) \leq \nu$.  This bound is tight, in
the sense that, for $p(1) < 2^{-\nu}$, one can always find a
probability mass function with $l(1) > \nu$.  Conversely, if $p(1)
\leq 1/(2^{\nu}-1)$, there is an optimal code with $l(1) \geq
\nu$, and this bound is also tight.
\label{mmprl1}
\end{theorem}

\begin{proof}
Suppose $p(1) \geq 2^{-\nu}$ and $l(1) \geq 1+\nu$.  Then
$R_\opt^*(p) = R^*(\boldl,p) \geq l(1) + \lg p(1) \geq 1$,
contradicting the simple bounds of (\ref{mmprbounds}).  Thus $l(1)
\leq \nu$.

For tightness of the bound, suppose $p(1) \in
(2^{-\nu-1},2^{-\nu})$ and consider $n=2^{\nu+1}$ and
$$p = \left(p(1), \underbrace{2^{-\nu-1}, \ldots, 2^{-\nu-1}}_{n-2},
2^{-\nu}-p(1)\right).$$ If $l(1) \leq \nu$, then, by the Kraft
inequality, one of $l(2)$ through $l(n-1)$ must exceed $\nu$.
However, this contradicts the simple bounds of (\ref{mmprbounds}).
For $p(1) = 2^{-\nu-1}$, a uniform distribution results in $l(1) =
\nu+1$.  Thus, since these two results hold for any $\nu$, this
extends to all $p(1)<2^{-\nu-1}$, and this bound is tight.

Suppose $p(1) \leq 1/(2^{\nu}-1)$ and consider an optimal length
distribution with $l(1) < \nu$.  Consider the weights of the nodes of
the corresponding code tree at level $l(1)$.  One of these weights is
$p(1)$, while the rest are known to sum to a number no less than
$1-p(1)$.  Thus there is one node of at least weight
$$\frac{1-p(1)}{2^{l(1)}-1} \geq \frac{1-p(1)}{2^{l(1)}-2^{l(1)+1-\nu}}$$
and thus, taking the logarithm and adding $l(1)$ to the right-hand side,
$$R^*(\boldl,p) \geq \nu-1 + \lg \frac{1-p(1)}{2^{\nu-1}-1}.$$
Note that $l(1)+1+\lg p(1) \leq \nu + \lg p(1) \leq \nu-1 + \lg
((1-p(1))/(2^{\nu-1}-1))$, a direct consequence of $p(1) \leq
1/(2^{\nu}-1)$.  Thus, if we replace this code with one for which
$l(1) = \nu$, the code is still optimal.  The tightness of the
bound is easily seen by applying Lemma~\ref{complete} to distributions
of the form
$$p = \left(p(1),\underbrace{\frac{1-p(1)}{2^\nu-2},
\ldots,\frac{1-p(1)}{2^\nu-2}}_{2^\nu-2}\right)$$ for $p(1)
\in (1/(2^{\nu}-1),1/2^{\nu-1})$.  This results in $l(1) = \nu-1$
and thus $R_\opt^*(p) = \nu + \lg (1-p(1)) - \lg (2^\nu-2)$,
which no code with $l(1) > \nu-1$ could achieve.
\end{proof}

In particular, if $p(1) \geq 0.5$, $l(1)=1$, while if $l(1) \leq 1/3$,
there is an optimal code with $l(1) > 1$.

We now briefly address the $d$\textsuperscript{th} exponential redundancy problem.
Recall that this is the minimization of
$$R^d(p,\boldl) \definedas \frac{1}{d} \lg \sum_{i \in \X} p(i)^{1+d}
2^{dl(i)}.$$ This can be rewritten as
$$R^d(p,\boldl) = \frac{1}{d} \lg \sum_{i \in \X} p(i) 2^{d(l(i)+\lg
p(i))}.$$ A straightforward application of Lyapunov's inequality for
moments yields $R^c(p,\boldl) \leq R^d(p,\boldl)$ for $c \leq d$,
which, taking limits to $0$ and $\infty$, results in
$$0 \leq \bar{R}(p,\boldl) \leq R^d(p,\boldl) \leq R^*(p,\boldl) < 1$$
for any valid $p$, $d > 0$, and $\boldl$, resulting in an extension of
(\ref{mmprbounds}), $$0 \leq \bar{R}_\opt(p) \leq R_\opt^d(p) \leq
R_\opt^*(p) < 1$$ where $R_\opt^d(p)$ is the optimal
$d$\textsuperscript{th} exponential redundancy, an improvement on the
bounds found in \cite{Baer05}.  This implies that this problem can be
bounded in terms of the most likely symbol using the upper bounds of
Theorem~\ref{mmprbetter} and the lower bounds of average redundancy
(Huffman) coding\cite{MoAb}:
$$\bar{R}_\opt \geq \xi - (1-p(1))\lg(2^\xi-1) - H(p(1), 1-p(1))$$
where $$\xi = \left\lceil \lg
\frac{1-2^\frac{1}{p(1)-1}}{1-2^\frac{p(1)}{p(1)-1}} \right\rceil$$
for $p(1) \in (0,1)$ (and, recall, $H(x) = - \sum_i x(i) \lg x(i)$).

\ifx \cyr \undefined \let \cyr = \relax \fi

\end{document}